# A note on retrodiction and machine evolution

Gustavo Caetano-Anollés
Evolutionary Bioinformatics Laboratory, Department of Crop Sciences and C. R. Woese Institute for Genomic Biology, University of Illinois, Urbana, Illinois, USA . Correspondence: gca@illinois.edu

**Abstract.** Biomolecular communication demands that interactions between parts of a molecular system act as scaffolds for message transmission. It also requires an evolving and organized system of signs – a communicative agency – for creating and transmitting meaning. Here I explore the need to dissect biomolecular communication with retrodiction approaches that make claims about the past given information that is available in the present. While the passage of time restricts the explanatory power of retrodiction, the use of molecular structure in biology offsets information erosion. This allows description of the gradual evolutionary rise of structural and functional innovations in RNA and proteins. The resulting chronologies can also describe the gradual rise of molecular machines of increasing complexity and computation capabilities. For example, the accretion of rRNA substructures and ribosomal proteins can be traced in time and placed within a geological timescale. Phylogenetic, algorithmic and theoretical-inspired accretion models can be reconciled into a congruent evolutionary model. Remarkably, the time of origin of enzymes, functional RNA, non-ribosomal peptide synthetase (NRPS) complexes, and ribosomes suggest they gradually climbed Chomsky's hierarchy of formal grammars, supporting the gradual complexification of machines and communication in molecular biology. Future retrodiction approaches and in-depth exploration of theoretical models of computation will need to confirm such evolutionary progression.

Retrodiction makes claims about the past given information that is available in the present. In contrast, prediction makes claims about the future based on information in the past. The two are intimately related. Travelling back in time or into the future (metaphorically) involves either a reverse-time or forward-time representation, which in stochastic dynamical systems bares on the amount of information (excess entropy) that is stored in the present.[1] In natural languages, scaling laws for excess entropy decay with the inverse square root of the number of non-terminal symbols in grammars, with symbols distributing according to the Heap's law and reflecting economy of scales and complexity accumulation in non-ergodic organization.[2,3] A time-symmetric representation can therefore unify the retrodiction-prediction paradigm, compressing it into an extant representation. In other words, the present holds enough information to retrodict and predict, a fact that is well known to evolutionary biologists.

Retrodiction is part of the 'ideographic' scientific method that focuses on process and history.[4] Together with the 'nomothetic' method that seeks to explain the present with universal (often mechanistic) statements, the ideographic method searches for irreducible and unrepeatable occurrences in the history of life. This evolutionary 'memory' is used to develop explanatory power. Retrodiction is at the interface of computational and experimental biology, taking full advantage of molecular evolution and systems and synthetic biology approaches. Computational biology methodology, for example, builds trees with or without reticulations (phylogenies) from both data and models of evolutionary change, which are often rooted by 'pulling down' a branch holding the ancestor.[4] These rooted trees can be scaled to time by converting them into 'timetrees'.[5] Computational approaches are not simply theoretical. Instead, they analyze contemporaneous data gathered from laboratory experimentation (e.g. genomic sequences, functional annotations) with models and algorithmic implementations that are also informed by biological knowledge acquired experimentally.[6] Computational phylogenetic approaches sometimes interface more directly with experimental science. For example, ancestors reconstructed by tracing evolutionary changes in the branches of phylogenetic trees, a procedure known as 'character state reconstruction' (CSR), can be used to 'resurrect' molecules which can then be analyzed with the tools of biochemistry and structural biology.[7] Resurrection can take many other forms, including extracting ancient nucleic acids from amber, ice cores, or revived organisms and viruses stored in the laboratory or preserved in permafrost. For example, a wide range of eukaryotic viruses (including giant viruses) have been recently revived from ancient Siberian permafrost, often from stomach or intestinal contents of fossil remains of woolly mammoth or Siberian wolf.[8] Their genomes have been sequenced and compared to those of extant viruses in phylogenomic reconstructions. In some particular cases, evolution can be traced in real time and the phylogenetic reconstruction accuracy of algorithms confirmed by experimentation. For example, SARS-CoV-2 viral variants are being collected, preserved, their genomes sequenced, and genomic data made freely available to the public thanks to a worldwide community effort (gisaid.org).[9] Exhaustive phylogenomic reconstructions describing viral evolution are being analyzed in real time as the COVID-19 pandemic continues to unfold worldwide.[10] For example, a recent analysis of ~12 million SARS-CoV-2 genomic sequences uncovered ~20,000 unique mutations, 22 haplotypes defining Variants of Concern Alpha, Delta and Omicron, and networks of protein interactions associated with seasonal effects.[11] Tracing the impact of mutations on protein structure with the powerful AlphaFold2 *ab initio* modeling pipeline complemented these studies by showcasing evolutionary changes in the structure and function of the viral proteins. Here, computational deep learning methods of structural prediction mimic reality with accuracies comparable to that of crystallographic methods.[12] Similarly, natural language processing applied to the analysis of proteomes and genomes predicts the identity of residues given their context.[13] These analyses can even anticipate phylogenetic trajectories.[14,15] The schism of theory and experimentation vanishes in these computational investigations, showing the increasing power of the retrodiction-prediction duality in the age of artificial intelligence.

The passage of time, however, restricts explanatory power and makes it increasingly difficult to find evolutionary memory embedded in extant data. Evolution modeled as a regular Markov chain of states reveals that more numerous and longer time steps decrease the probability of descendants having the states of ancestors.[16,17] First, Markov chains lose information exponentially with the passage of time in a process that is only countered by drift or balancing selection. Second, the 'data processing inequality' of causal chains, which is an internal property of the Markov chain and is independent of temporal information loss, ensures that information content of a signal cannot be increased via local operators. In other words, given a Markov chain $X \to Y \to Z$, mutual information of deterministic or random processing must comply with $I(X;Y) \geq I(X;Z)$. Third, the mapping of $X$, $Y$ and $Z$ may not be one-to-one. Instead, ancient ancestors may represent ancient relatives and many past events could map to the present. The more entangled the path from the past to the present, the more historical is the retrodiction enterprise and the more difficult its dissection with statistical approaches such as maximum likelihood or Bayesian analysis. Thus, time erodes information in the present that can explain the past.

One way to offset this information erosion is to focus on molecular structure, which is orders of magnitude more evolutionarily conserved that sequence.[18] In Fig. 1, I provide two examples of structural retrodiction, one involving the ribosome[19] and the other the proteome.[20] While information in RNA structure can improve sequence alignments[21] or be used directly to build standard phylogenetic trees,[22,23] the first studies of molecular growth over evolutionary time (accretion) made use of CSRs along branches of a tree of life generated from ribosomal RNA (rRNA)[24] or reconstructed trees of rRNA substructures describing ribosomal growth.[25] These phylogenetic strategies allow to study molecular accretion in different RNA molecules, including RNA of ancient origin such as transfer RNA (tRNA), 5S rRNA, RNase P RNA, SINE RNA and rRNA (reviewed in ref. [26]). Operationally, geometrical features (e.g., length of single-stranded or double-stranded RNA segments) or statistical features (e.g. Shannon entropy of the base-pairing probability matrix) of substructures were coded into linearly ordered multistate characters in data matrices for phylogenetic analysis. Trees derived from geometrical and statistical characters were congruent, mutually supporting the retrodiction exercise. In parallel, the first study of molecular growth of the protein world reconstructed trees of protein structural domains from a molecular census of SCOP fold structures.[27] Later retrodictions also used CSR approaches and extended evolutionary exploration to SCOP domain superfamilies and families, CATH architectures, topologies and homologous superfamilies, and proteins defined as combinations of domain structures (reviewed in ref.[28]). Remarkably, a molecular clock of folds[29], which has been extended to superfamilies and families,[28] was able to link structural and geological timescales. This clock allowed to assign ages in billions of years (Gy) to structural domains, confirming rooting and evolutionary statements derived from phylogenomic reconstruction. In most of these studies, trees of RNA substructures and protein domains were rooted with the Lundberg method without resorting to a molecular clock model,[4] using either the 'standard' implementation, which invokes Weston's generality criterion by either optimizing ancestral-derived homology relationships in nested patterns along branches of the trees or by defining a maximum or minimum state ancestor according to considerations of RNA conformational stability or abundance of domains in proteomes. Both implementations have been shown to produce topologically isomorphic rooted trees, mutually validating the rooting approaches [e.g., refs.[30]].

If molecular machines are vehicles of communication in biology, their grammars should climb up Chomsky's hierarchy[31,32] when their structural makeup gradually increase in complexity. The regular grammars of metabolic enzymes, would be followed in evolution by context-free grammars of functional RNA, context-sensitive grammars of non-ribosomal peptide synthetase (NRPS) complexes, and unrestricted grammars of the ribosomal protein biosynthetic machinery, in that specific order. Phylogenomic analysis of protein domains provide considerable support to this progression, beginning with an analysis of the metabolic origins of translation, proteins, and protein biosynthesis.[33,34] In these studies, chronologies revealed the gradual build-up of domain repertoires associated with metabolism, followed by tRNA-interacting translation factors and aminoacyl-tRNA synthetases (aaRSs), NRPS modules that synthetize small peptides in assembly lines, and finally, ribosomal proteins (r-proteins) needed for ribosomal functionality. Tracing the origin of cofactors in these chronologies showed the ancestral nature of ATP/ADP (as previously suggested[35]), immediately followed by the ubiquitous NAD family of redox cofactors.[34] Interactions with primordial tRNA and rRNA ligands appeared later. The analysis of the evolutionary origins of the ribosome[19] illustrated with the evolutionary timeline of accretion of both rRNA and ribosomal proteins (r-protein) of Fig. 1B reveals co-evolution of rRNA and r-proteins and co-evolution of the two major subunits of the ribosome. Substructures needed for decoding, helicase functions, and ribosomal mechanics (e.g., ratchet, central protuberance), preceded the peptidyl transferase center (PTC) responsible for protein biosynthesis, suggesting the most primordial elements were probably associated with RNA processing (perhaps a primitive replication apparatus) driven by regular and context-free grammars. These results challenge the ancient 'RNA world' narrative that prevails in origin of life research.[36] Instead, they align with significant and fundamentally biochemical and resurrective evidence that suggests the genetic code originated in an aaRS 'urzyme' protein biology.[37]

The validity of the ribosomal phylogenetic accretion model has been tested against algorithmic and theoretical (nomothetic) implementations. An algorithm of rRNA accretion based on A-minor interactions and periphery-core ribosomal dismantling of the large subunit rRNA[38] was compatible with the history of A-minor interactions of the phylogenetic model[19] once equally-likely terminal disassembly steps were not artificially forced towards an origin in the PTC and translocation structures were not dismantled in first steps.[19,26] An alternative algorithmic implementation grew molecules outwards (onion-like) from the PTC, which was assumed to represent the most ancient substructure of the molecule.[39,40] The algorithm inserted "branch" helices onto preexisting, coaxially-stacked, "trunk" helices , leaving in the process "insertion fingerprint" constrictions in their junctions (a natural property of RNA junctions). While the algorithm demanded absence of 'trunk-to-branch' roadblocks to outward growth, we identified at least 17 roadblocks in the small and large ribosomal subunits creating 19 possible ribosomal origins.[41] Remarkably, accounting for them added to the branch-to-trunk insertion sequence an additional older phase holding translocation structures, again reconciling the phylogenetic[19] and algorithmic[39] models and revealing commonalities: burst-like appearance of the PTC region, construction of an exit tunnel by gradual accumulation of structural layers, and 3-dimensional layering patterns radiating from a central core.[26] More recently, properties of *in silico*-designed 'RNA ring' constructs mimicking ancestral biomolecules (likely ancient tRNAs) tested whether ring substructures accreted to form rRNA.[42,43] Remarkably, when times of origin of rRNA substructures of the phylogenetic and algorithmic models were compared against those of theoretical minimal RNA rings, the ages of the phylogenetic model showed a better match.[41] Thus, approaches converge, all of them supporting the process of ribosomal accretion illustrated in Fig. 1B.

The late evolutionary arrival of processive protein biosynthesis is an expected consequence of the gradual complexification of communication networks, starting with regular grammars in the processing of metabolic substrates by primordial ribozymes and proteinogenic enzymes and ending with a ribosomal universal Turing machine[26] enabled by genetic encoding in an aaRS duality.[37] Besides coordinating the initially loosely connected communication networks, the ribosomal computational capabilities made uniform the initial diversity of molecular languages (e.g., by sieving thousands of amino acid monomers of the type used by modern NRPSs) and the ability to remember actions that occurred in the past (a property of Turing machines). In fact, the ribosome may well be a Turing machine with modified 'temporal memory' of the type found in the co-called 'Stateless Bounded Temporal Memory' Turing machine that lacks multiple control states,[44] or the unconventional 'stateless' Turing machine with mobile heads that move in single steps but encompass three placeholders for symbols in tapes,[45] both of which are Turing complete. New computational architectures would avoid the burdensome memorization of control states that existed in the distant past fueled by the initial chemical diversity of early Earth and at the same time help the emerging machine learn how to more efficiently generate end-directed behaviors.

In the future, retrodiction approaches and in-depth exploration of theoretical models of computation will likely confirm the evolutionary complexification of machines and communication. Future resurrection experiments could also bring putative ancestral ribosomes back to life, with which to test their likelihoods, functionalities, and computational capabilities. This brand-new world of synthetic biology will benefit from retrodiction and deep learning predictive science.

**References**

1. Ellison, C.J., Mahoney, J.R., & Crutchfield, J.P. (2009). Prediction, retrodiction, and the amount of information stored in the present. *Journal of Statistical Physics*, *136*, 1005-1034.
2. Ebeling, W., & Pöschel, T. (1994). Entropy and long-range correlations in literary English. *Europhysics Letters*, *26*(4), 241.
3. Debowski, L. (2011). On the vocabulary of grammar-based codes and the logical consistency of texts. IEEE *Transactions on Information Theory*, *57*(7), 4589-4599.
4. Caetano-Anollés, G., Nasir, A., Kim, K.M., & Caetano-Anollés, D. (2018). Rooting phylogenies and the Tree of Life while minimizing *Ad Hoc* and auxiliary assumptions. Evolutionary Bioinformatics, 14, 1176934318805101.
5. Kumar, S., Suleski, M., Craig, J., Kasprowicz, A., Sanderford, M., Li, M., Stecher, G., & Hedges, S.B. (2022) TimeTree5: An expanded resource for species divergence times. *Molecular Biology and Evolution*, *39*(8), msac174.
6. Haber, M., & Velasco, J. (2022). Phylogenetic inference., In E.N. Zalta & U. Nodelman (Eds.), *The Stanford Encyclopedia of Philosophy*. https://plato.stanford.edu/archives/fall2022/entries/phylogenetic-inference

7. Zaucha, J., & Heddle, J.G. (2017). Resurrecting the dead (molecules). *Computational and Structural Biotechnology Journal, 15*, 351-358.
8. Alempic, J.-M., Lartigue, A., Goncharov, A.E., et al. (2023). An update on eukaryotic viruses revived from ancient permafrost. *Viruses*, *15*, 564.
9. Elbe, S., & Buckland-Merrett, G. (2017). Data, disease and diplomacy: GISAID's innovative contribution to global health. *Global Challenges*, *1*, 33–46.
10. Talenti, A., Hodcroft, E.B., & Robertson, D.L. (2022). The evolution and biology of SARS-CoV-2 variants. *Cold Spring Harbor Perspectives in Medicine*, *12*, a041390.
11. Tomaszewski, T., Ali, M.A., Caetano-Anollés, K., & Caetano-Anollés, G. (2023). Seasonal effects decouple SARS-CoV-2 haplotypes worldwide. *F1000Research*, *12*, 267.
12. Jumper, J., Evans, R., Pritzel, A., et al. (2021). Highly accurate protein structure prediction with AlphaFold. *Nature. 596*, 583–589.
13. Asgari, E., & Mofrad, M.R.K. (2015). Continuous distributed representation of biological sequences for deep proteomics and genomics. *PLoS One, 10*(11), e0141287.
14. Lupo, U., Sgarbossa, D., & Bitbol, A.F. (2022). Protein language models trained on multiple sequence alignments learn phylogenetic relationships. *Nature Communications*, *13*(1), 6298
15. Hie, B.L., Yang, K.K., & Kim, P.S. (2022). Evolutionary velocity with protein language models predicts evolutionary dynamics of diverse proteins. *Cell Systems*, *13*(4), 274–285.
16. Sober, E., & Steel, M. (2014). Time and knowability in evolutionary processes. *Philosophy of Science*, *81*, 537-557.
17. Sober, E., & Steel, M. (2015). Similarities as evidence for common ancestry: A likelihood epistemology. *British Journal of Philosophy of Science*, *68*, 617-638.
18. Caetano-Anollés, G., & Nasir, A. (2012). Benefits of using molecular structure and abundance in phylogenomic analysis. *Frontiers in Genetics*, *3*, 172.
19. Harish, A., & Caetano-Anollés, G. (2012). Ribosomal history reveals origins of modern protein synthesis. *PLoS One*, *7*(3), e32776.
20. Aziz, M.F., & Caetano-Anollés, G. (2021). Evolution of networks of protein domain organization. Scientific Reports, 11, 12075.
21. Higgs, P.G. (2000). RNA secondary structure: physical and computational aspects. *Quarterly Reviews of Biophysics*, *33*, 199-253.
22. Billoud, B., Guerrucci, M-A., Masselot, M., & Deutsch, J.S. (2000). Cirripede phylogeny using a novel approach: Molecular morphometrics. *Molecular Biology and Evolution*, *17*(10), 1435-1445.
23. Collins, L.J., Moulton, V., & Penny, D. (2000). Use of RNA secondary structure for studying the evolution of RNase P and RNase MRP. *Journal of Molecular Evolution*, *51*, 194–204.
24. Caetano-Anollés, G. (2002). Evolved RNA secondary structure and the rooting of the universal tree of life. *Journal of Molecular Evolution*, *54*, 333–345.
25. Caetano-Anollés, G. (2002). Tracing the evolution of RNA structure in ribosomes. *Nucleic Acids Research, 30*, 2575–2587.
26. Caetano-Anollés, G., & Caetano-Anollés, D. (2015). Computing the origin and evolution of the ribosome from its structure — Uncovering processes of macromolecular accretion benefiting synthetic biology. *Computational and Structural Biotechnology Journal, 13*, 427–447.
27. Caetano-Anollés, G., & Caetano-Anollés, D. (2003). An evolutionarily structured universe of protein architecture. *Genome Research*, *13*, 1563–1571.
28. Caetano-Anollés, G., Aziz, M.F., Mughal, F., & Caetano-Anollés, D. (2021) Tracing protein and proteome history with chronologies and networks: folding recapitulates evolution. *Expert Reviews in Proteomics, 18(10)*, 863-880.
29. Wang, M., Jiang, Y.-Y., Kim, K.M., Qu, G., Ji, H.-F., Mittenthal, J.E., Zhang, H.-Y., & Caetano-Anollés, G. (2011). A universal molecular clock of protein folds and its power in tracing the early history of aerobic metabolism and planet oxygenation. *Molecular Biology and Evolution*, *28*(1), 567–582.
30. Caetano-Anollés, D., Nasir, A., Kim, K.M., & Caetano-Anollés, G. (2019). Testing empirical support for evolutionary models that root the Tree of Life. *Journal of Molecular Evolution*, *87*, 131-142.
31. Chomsky, N. (1956). Three models for the description of language. *IRE Transanctions on Information Theory*, *2*(3), 113-124.

32. Chomsky, N. (1959). On certain formal properties of grammars. *Information and Control, 2*, 137-167.
33. Caetano-Anollés, D., Kim, K.M., Mittenthal, J.E., & Caetano-Anollés, G. (2011). Proteome evolution and the metabolic origins of translation and cellular life. *Journal of Molecular Evolution*, *72*, 14-33.
34. Caetano-Anollés, G., Kim, K.M., & Caetano-Anollés, D. (2012). The phylogenomic roots of modern biochemistry: Origins of proteins, cofactors, and protein biosynthesis. *Journal of Molecular Evolution*, *74*, 1-34.
35. Ji, H.F., Kong, D.X., Shen, L., Chen, L.L., Ma, B.G., & Zhang, H.-Y. (2007). Distribution patterns of small molecule ligands in the protein universe and implications for origins of life and drug discovery. *Genome Biology, 8*, R176
36. Robertson, M.P., & Joyce, G.P. (2012). The origins of the RNA world. *Cold Spring Harbor Perspectives in Biology, 4*, a004608.
37. Carter, C.W., & Wills, P.R. (2021). The roots of genetic coding in aminoacyl-tRNA synthetase duality. *Annual Review of Biochemistry*, *90*, 349-373.
38. Bokov, K., Steinberg, S.V. (2009). A hierarchical model for evolution of 23S ribosomal RNA. *Nature*, *457*(7232), 977–980.
39. Petrov, A.S., Bernier, C.R., Hsiao, C., Norris, A.M., Kovacs, N.A., Waterbury, C.C., Stepanov, V.G., Harvey, S.G., Fox, G.E., Wartell, R.M., Hud, N.V., & Williams, L.D. (2014). Evolution of the ribosome at atomic resolution. *Proceedings of the National Academy of Science of the USA*, *111*, 10251–10256.
40. Petrov, A. S., Gulen, B., Norris, A.M., Kovacs, N.A., Bernier, C.R., Lanier, K.A., Fox, G.E., Harvey, S.C., Wartell, R.M., Hudd, N.V., & Williams, L.D. (2015). History of the ribosome and the origin of translation. *Proceedings of the National Academy of Science of the USA*, *112*, 15396–15401.
41. Caetano-Anollés, D., & Caetano-Anollés, G. (2017). Commentary: History of the ribosome and the origin of translation. *Frontiers in Molecular Bioscience*, *3*, 87.
42. Demongeot, J., & Seligmann, H. (2020). Accretion history of large ribosomal subunits deduced from theoretical minimal RNA is congruent with histories derived from phylogenetic and structural methods. *Gene*, *738*, 144436.
43. Demongeot, J., & Seligmann, H. (2020). Comparison between small ribosomal RNA and theoretical minimal RNA ring secondary structures confirm phylogenetic and structural histories. *Scientific Reports*, *10*, 7693.
44. Cantone, D., & Cristofaro, S. (2016). A variant of Turing machines with no controls states and its connection to bounded temporal memory. In V. Bilo & A. Caruso (Eds.), *Proceedings of the 17th Italian Conference on Theoretical Computer Science*, CEUR Workshop Proceedings, Volume 1720, pp. 36-48.
45. Arulanansham, J.J. (2007). Unconventional 'stateless' Turing-like machines. In S.G. Akl, C.S. Calude, M.J. Dineen, G. Rozenberg, & H.T. Wareham (Eds.), *Proceedings of the 6th International Conference on Unconventional Computation*, Lecture Notes in Computer Science, Volume 4618, pp. 55-61.

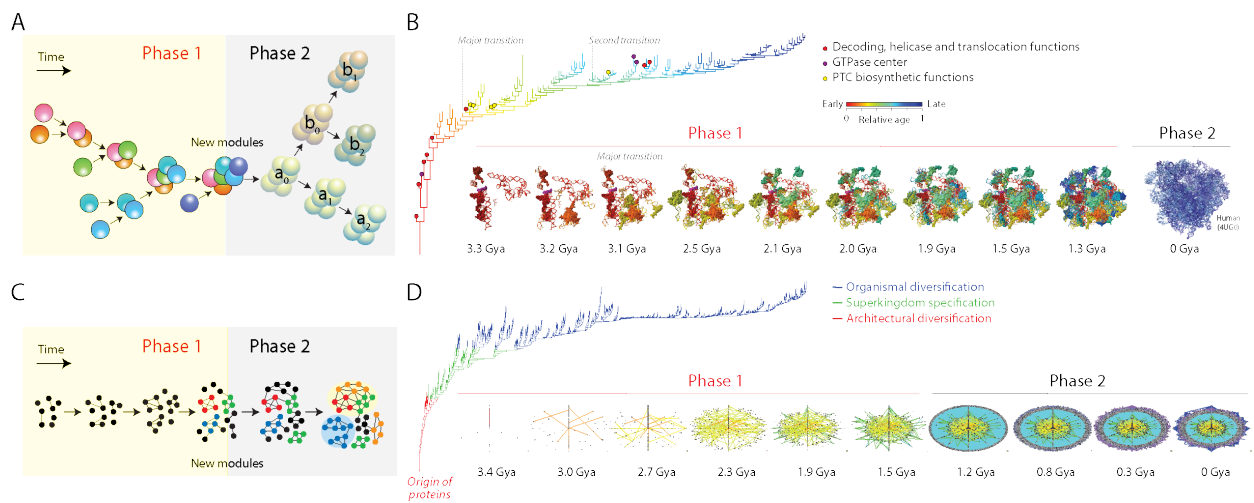

**FIGURE 1** The biphasic paradigm explained with time series and examples. (A) A tree paradigm describe accumulating events of accretion and diversification. Individual parts (spheres) come together to form a module (Phase 1), which then diversifies in sequential (modules $a_0$, $a_1$ and $a_2$) or branching (modules $b_0$, $b_1$ and $b_2$) manner in absence or presence of other evolving modules (Phase 2). (B) The biphasic history of the ribosome is showcased by an evolutionary timeline of ribosomal RNA (rRNA) and proteins (r-proteins) inferred directly from phylogenomic data. During an initial phase (Phase 1), helical structures of rRNA and r-proteins accreted to form a universal ribosomal core. The second phase of ribosomal evolution (Phase 2) started 1.3 billions of years ago (Gya) (or earlier) when the universal core diversified alongside with evolving organismal lineages. The phylogenomic tree describes the accretion of rRNA helical stems and is colored according to time of origin (relative age).[19] Every new branch reflects the addition of a new part to the whole. The first RNA structures to accrete include the head and ratchet, the central protuberance, and stalks, which are involved in ribosomal dynamics. Early structures are also involved in energetics, decoding, helicase activity, and translocation. The peptidyl transferase center (PTC) that is responsible for protein biosynthesis accretes later in time (in yellow), whereas RNA helices gradually gained interaction with r-proteins to form a processivity core 2.8 to 3.1 Gya at a time when a crucial "major transition" in ribosomal evolution brought small and large subunits together by formation of intersubunit bridges. A molecular clock of folds linked structural and geological timescales.[29] (C) A network paradigm describes the two evolutionary phases with a time series of an evolving graph in which nodes and links describe parts and interactions in an evolving biological system, respectively. The rise of hierarchical modularity during Phase 1 produces highly connected communities (subnetworks), which become modules when their interactions stabilize. In Phase 2, modules coalesce into higher level network substructures. (D) A time series of networks describes the evolution of protein domain organization. Snapshots of evolving networks of the compositional CX type[20] taken at regular intervals are described in radial format. Nodes of the networks represent architectures defined at SCOP superfamily level and color-coded arcs represent donor-acceptor recruitments and flow of time from ancient to recent nodes. A most-parsimonious phylogenomic tree of protein architectures shows the three epochs of the protein world.